\documentclass[12pt]{article}
\usepackage{ascmac}
\usepackage{ulem}
\usepackage{latexsym}
\usepackage{graphicx}
\usepackage{amsmath}
\usepackage[top=3.5cm,bottom=3.5cm,left=2cm,right=2cm]{geometry}
\usepackage{pifont}
\usepackage{cancel}             
\usepackage{amsmath,amssymb}

\newcommand{\h}{\hspace}
\newcommand{\be}{\begin{equation}}
\newcommand{\e}{\end{equation}}

\begin{document}

\title{
\vbox{ 
\baselineskip 14pt
\hfill \hbox{\normalsize KUNS-2528}
} \vskip 1cm
\bf \Large Multiple Point Principle of the  Standard Model with Scalar Singlet Dark Matter and Right Handed Neutrinos\vskip 0.5cm
}
\author{
Kiyoharu~Kawana\thanks{E-mail: \tt kiyokawa@gauge.scphys.kyoto-u.ac.jp}
\bigskip\\
\it \normalsize
 Department of Physics, Kyoto University, Kyoto 606-8502, Japan\\
\smallskip
}
\date{\today}

\maketitle

\abstract{\noindent\normalsize
We consider the multiple point principle (MPP) of the Standard Model (SM) with the scalar singlet Dark Matter (DM) and three heavy right-handed neutrinos at the scale where the beta function $\beta_{\lambda}$ of the effective Higgs self coupling $\lambda_{\text{eff}}$ becomes zero. We make the two-loop analysis and  find that the top quark mass $M_{t}$ and the Higgs portal coupling $\kappa$ are strongly related each other. One of the good points in this model is that the larger $M_{t}\h{1mm}(\gtrsim 171\text{GeV})$ is allowed. This fact is consistent with the recent experimental value \cite{ATLAS:2014wva} $M_{t}=173.34\pm0.76$ GeV, which corresponds to the DM mass $769\h{1mm}\text{GeV}\leq m_{\text{DM}}\leq 1053\h{1mm}\text{GeV}$. 

\newpage

\section{Introduction}
The discovery of the Higgs like particle and its mass \cite{Aad:2012tfa,Chatrchyan:2012ufa} is a very meaningful result for the Standard Model (SM). It suggests that the Higgs potential can be stable up to the Planck scale $M_{pl}$ and also that both of the Higgs self coupling $\lambda$ and its beta function $\beta_{\lambda}$ become very small around the Planck scale. This fact attracts much attention, and there are many works which try to find its physical meaning \cite{Froggatt:1995rt,Froggatt:2001pa,Nielsen:2012pu,Kawana:2015tka,Shaposhnikov:2009pv,Meissner:2007xv,Khoze:2014xha,Kawai:2011qb,Kawai:2013wwa,Hamada:2014ofa,Hamada:2014xra,Hamada:2013mya,Hamada:2014iga,Hamada:2014wna,Hamada:2014raa,Kawamura:2013kua,Meissner:2006zh,Haba:2014sia,Iso:2012jn,Hamada:2015ria}.

One of the interesting and meaningful studies is to consider how the physics beyond the SM affects such a criticality. For example, recently there has been a two loop analysis about the Higgs portal $Z_{2}$ scalar model \cite{Hamada:2014xka}. In this model, the SM singlet scalar is a Dark Matter (DM) candidate, and it is found that its mass can be predicted to be $400\text{GeV}<m_{DM}<470\text{GeV}$ from the requirement that $\lambda$ and $\beta_{\lambda}$ simultaneously become zero at $10^{17}$GeV, which is usually called the multiple point principle (MPP) \cite{Froggatt:1995rt,Froggatt:2001pa,Nielsen:2012pu,Kawana:2015tka}. 

In this paper, we study the MPP of the next minimal extension of the SM, namely, besides the Higgs portal $Z_{2}$ scalar, we include SM singlet  heavy right-handed neutrinos \cite{Haba:2014sia,Haba:2014zda,Haba:2014zja}. The MPP of this model at the (reduced) Planck scale $M_{pl}$ has already been investigated in \cite{Haba:2014sia}. There, by using the two-loop beta functions and the tree-level Higgs potential, they concluded that  $m_{DM}$ and the heavy Majorana mass $M_{R}$ of the right-handed neutrino should be
\begin{align}&8.5\h{1mm}(8.0)\times10^{2}\h{1mm}\text{GeV}\leq m_{DM}\leq1.4\h{1mm}(1.2)\times10^{3}\h{1mm}\text{GeV}, \\
&6.3\h{1mm}(5.5)\times10^{13}\h{1mm}\text{GeV}\leq M_{R}\leq1.6\h{1mm}(1.2)\times10^{14}\h{1mm}\text{GeV},\end{align}
within $172.6\h{1mm}\text{GeV}\leq M_{t}\leq 174.1\h{1mm}\text{GeV}$. The different points in this paper are as follows:
\begin{enumerate}\item We consider the MPP at the scale where $\beta_{\lambda}$ becomes zero. Namely, we do not fix the MPP scale at $M_{pl}$. As a result, the condition $\beta_{\lambda}=0$ does not reduce the degrees of freedom of parameters.\\
\item In addition to the two-loop beta functions, we also calculate the one-loop effective potential.\\
\item We fix $M_{R}$ to $10^{13}$ GeV, and include the Yukawa coupling $Y_{R}$ between the $Z_{2}$ scalar and the right-handed neutrinos.
\end{enumerate}
Although, within the renormalizable Lagrangian, there are also two scalar couplings in this model (see Eq.(\ref{eq:lag1})), we focus on $\lambda$ (and $\beta_{\lambda}$) in this paper \footnote{It is difficult to realize the MPP of the other scalar couplings simultaneously in addition to $\lambda$. This is discussed in \ref{app:MPP}.}. The existence of heavy right-handed neutrinos is naturally needed if we try to explain the light neutrino masses by the seesaw mechanism. Thus, this model is phenomenologically interesting because it can explain both of DM and the light neutrino masses. \\

This paper is organized as follows. In Section\ref{sec:SM}, we review the MPP of the pure SM for the later discussion. In Section\ref{sec:two}, we give the two-loop analysis of the SM with the scalar singlet DM and three right handed neutrinos. In Section\ref{sec:summary}, the summary is  given.

\section{Preliminary - Multiple Point Principle  of SM -}\label{sec:SM}
In the SM, the one loop effective potential in Landau gauge is given by
\be V_{\text{eff}}(\phi,\mu)=V_{\text{tree}}(\phi,\mu)+V^{\text{SM}}_{\text{1loop}}(\phi,\mu),\e
where
\be V_{\text{tree}}(\phi,\mu):=e^{4\Gamma(\phi)}\frac{\lambda(\mu)}{4}\phi^{4},\e
\begin{align} V_{\text{1loop}}(\phi)&:=e^{4\Gamma(\phi)}\Biggl\{-6\cdot\frac{M_{t}(\phi)^{4}}{64\pi^{2}}\left[\log\left(\frac{M_{t}^{2}(\phi)}{\mu^{2}}\right)-\frac{3}{2}+2\Gamma(\phi)\right] \nonumber\\
&+3\cdot\frac{M_{W}(\phi)^{4}}{64\pi^{2}}\left[\log\left(\frac{M_{W}^{2}(\phi)}{\mu^{2}}\right)-\frac{5}{6}+2\Gamma(\phi)\right]+3\cdot\frac{M_{Z}(\phi)^{4}}{64\pi^{2}}\left[\log\left(\frac{M_{Z}^{2}(\phi)}{\mu^{2}}\right)-\frac{5}{6}+2\Gamma(\phi)\right]\Biggl\},\end{align}
\be M_{t}(\phi)=\frac{y_{t}(\mu)}{\sqrt{2}}\phi\h{2mm},\h{2mm}M_{W}(\phi)=\frac{g_{2}(\mu)}{2}\phi\h{2mm},\h{2mm}M_{Z}=\frac{\sqrt{g_{2}^{2}(\mu)+g_{\text{Y}}^{2}(\mu)}}{2}\phi.\e
Here, $\mu$ is the renormalization scale, $\Gamma(\phi)$ is the wave function renormalization and $\lambda(\mu)$, $y_{t}(\mu)$, $g_{2}(\mu)$ and $g_{Y}(\mu)$ are the renormalized couplings\footnote{For the beta functions of the SM, see \cite{Hamada:2014xka,Buttazzo:2013uya,Hamada:2012bp} for example. Or we can reproduce them by using the results in \ref{app:beta}. }. By using those results, the effective Higgs self coupling $\lambda_{\text{eff}}(\phi,\mu)$ can be defined as
\be V_{\text{eff}}(\phi,\mu):=\frac{\lambda_{\text{eff}}(\phi,\mu)}{4}\phi^{4}.\e
To minimize the contribution of $V^{\text{SM}}_{\text{1loop}}(\phi,\mu)$, we put $\phi=\mu$ in the following discussion. 

\begin{figure}
\begin{minipage}{0.5\hsize}
\begin{center}
\includegraphics[width=8.2cm]{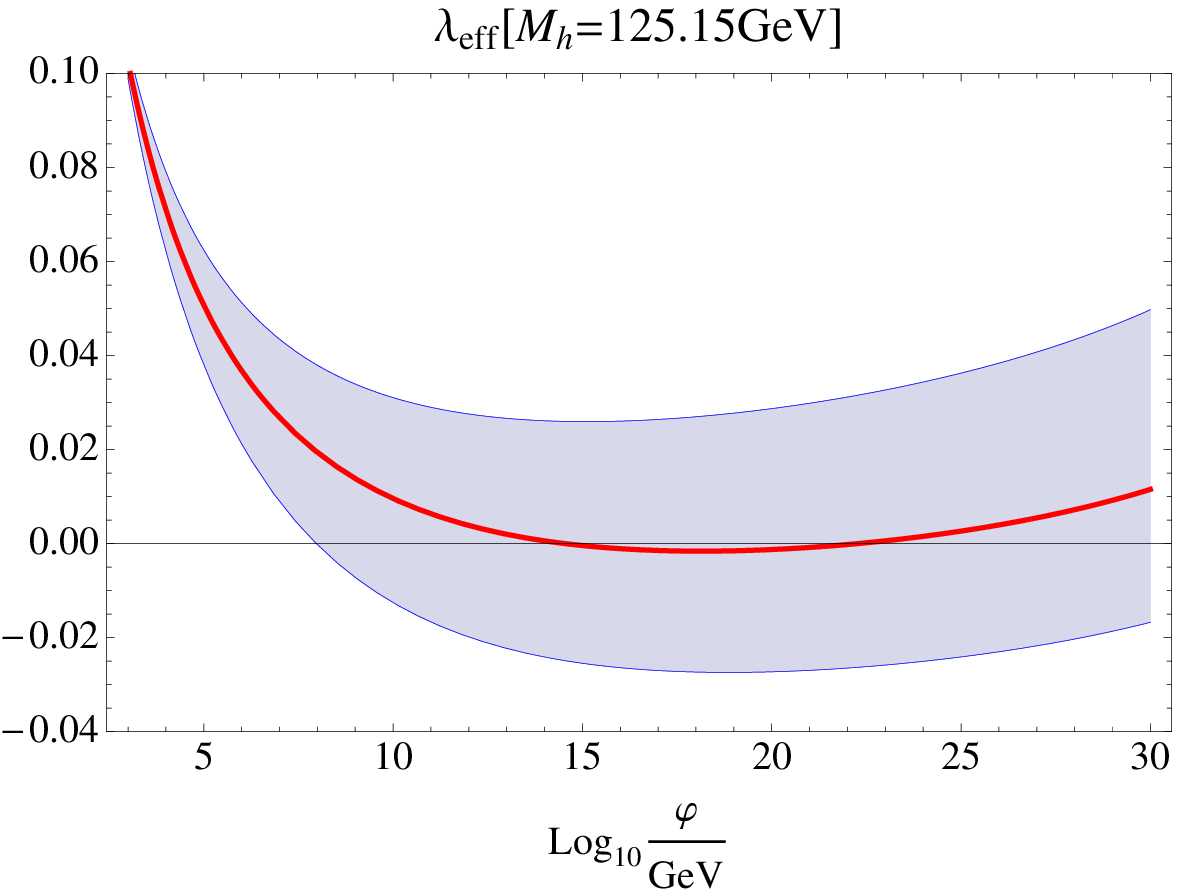}
\end{center}
\end{minipage}
\begin{minipage}{0.5\hsize}
\begin{center}
\includegraphics[width=8cm]{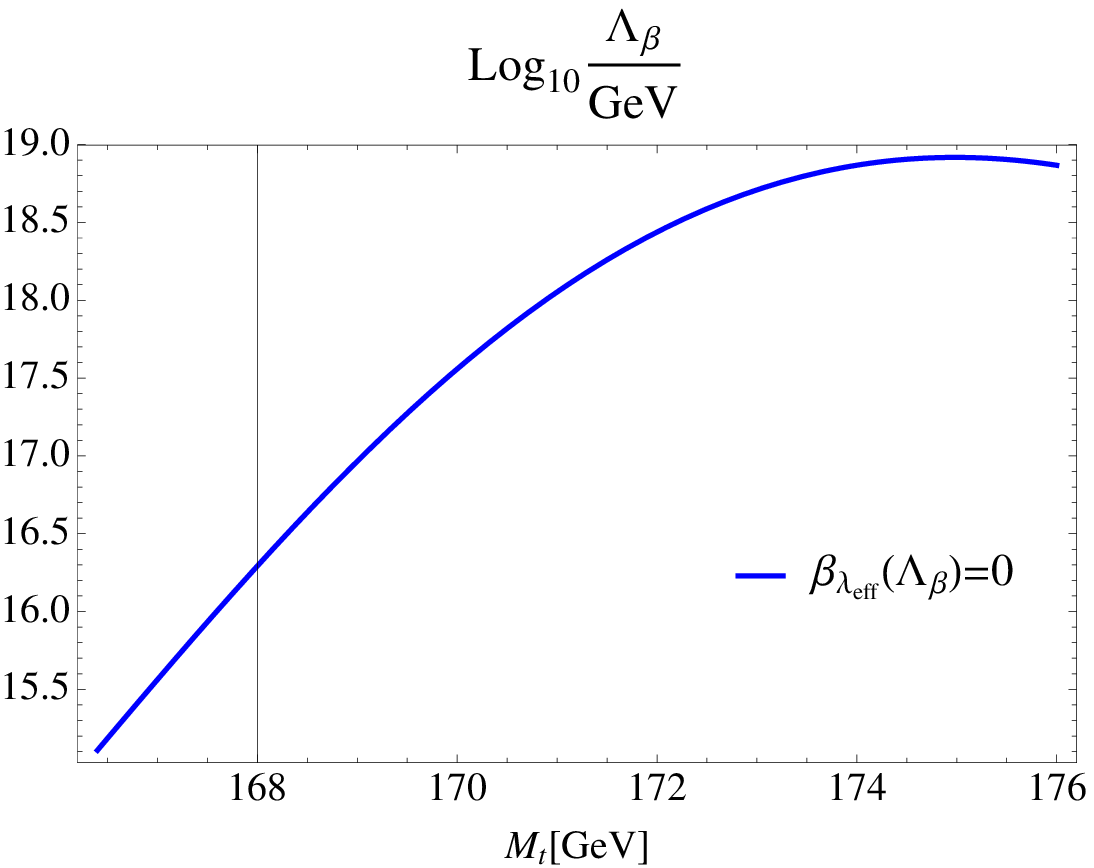}
\end{center}
\end{minipage}
\caption{Left panel shows the running effective Higgs self coupling $\lambda_{\text{eff}}$ as a function of the Higgs field $\phi$. The blue band corresponds to $95\%$ CL deviation of the top quark pole mass $M_{t}$. Right panel shows the scale $\Lambda_{\beta}$ where $\beta_{\lambda_{\text{eff}}}$ becomes zero as a function of $M_{t}$.} 
\label{fig:SM1}
\end{figure}

The left panel of Fig.1 shows $\lambda_{\text{eff}}(\phi)$ as a function of $\phi$. For the initial values, we have used the numerical results of \cite{Buttazzo:2013uya}, and the Higgs mass is fixed at
\be M_{h}=125.15\text{GeV}.\label{eq:higgs}\e
We use Eq.(\ref{eq:higgs}) as a typical value in the following discussion. The band corresponds $95\%$ CL deviation of the top quark pole mass $M_{t}$. For the 1$\sigma$ level, this is given by \cite{Moch:2014tta}
\be M_{t}=171.2\pm2.4\text{GeV}.\e
 If we assume that all the other parameters of the SM except for $M_{t}$ are fixed, we can find the scale $\Lambda_{\beta}$ where $\beta_{\lambda_{\text{eff}}}$ becomes zero as a function of $M_{t}$. Here, $\beta_{\lambda_{\text{eff}}}$ means
\be \beta_{\lambda_{\text{eff}}}(\phi):=\frac{d\lambda_{\text{eff}}(\phi)}{d\log\phi}.\e
The right panel of Fig.1 shows $\Lambda_{\beta}$ as a function of $M_{t}$. The MPP requires that $\lambda_{\text{eff}}(\Lambda_{\beta})$ should become zero, and predicts
\be M_{t}=170.9\text{GeV}.\e
This is the MPP of the pure SM. In the next section, we discuss the MPP of the SM with the scalar singlet DM and three right-handed neutrinos.

\section{MPP of the SM with Scalar Singlet Dark Matter and Right Handed Neutrinos}\label{sec:two}
We consider the following renormalizable Lagrangian:
\begin{align} {\cal{L}}={\cal{L}}_{\text{SM}}+\frac{1}{2}\partial_{\mu}&S\partial^{\mu}S-\frac{m_{\text{DM}}^{2}}{2}S^{2}-\frac{\kappa}{2}S^{2}H^{\dagger}H-\frac{\lambda_{\text{DM}}}{4!}S^{4}+\sum_{j=1}^{3}\bar{\nu}_{Rj}i\gamma^{\mu}\partial_{\mu}\nu_{Rj}\nonumber\\
&-\sum_{i,j}\left(y_{\nu ij}\bar{L}_{i}H^{\dagger}\nu_{Rj}+\text{h.c}\right)-\sum_{i,j}\left(M_{Rij}+\frac{Y_{Rij}}{\sqrt{2}}S\right)\bar{\nu}_{Ri}^{c}\nu_{Rj}.\label{eq:lag1}\end{align}
Here, $H$ is the Higgs field, $S$ is the SM singlet real scalar field, $m_{\text{DM}}$ is its mass, $\nu_{Ri}$ are right-handed neutrinos, $M_{Rij}$ are their Majorana masses,  and $(Y_{Rij},y_{\nu ij})$ are the Yukawa couplings. For simplicity, we assume that $M_{Rij}$, $Y_{Rij}$ and $y_{\nu ij}$ are diagonalized, and also that they are equal respectively for the three generations. In this case, Eq.(\ref{eq:lag1}) becomes
\begin{align} {\cal{L}}={\cal{L}}_{\text{SM}}+\frac{1}{2}\partial_{\mu}S\partial^{\mu}S-&\frac{m_{\text{DM}}^{2}}{2}S^{2}-\frac{\kappa}{2}S^{2}H^{\dagger}H-\frac{\lambda_{\text{DM}}}{4!}S^{4}+\sum_{i=1}^{3}\bar{\nu}_{Ri}i\gamma^{\mu}\partial_{\mu}\nu_{Ri}\nonumber\\
&-y_{\nu}\sum_{i=1}^{3}\left(\bar{L}_{i}H^{\dagger}\nu_{Ri}+\text{h.c}\right)-\sum_{i=1}^{3}\left(M_{R}+\frac{Y_{R}}{\sqrt{2}}S\right)\bar{\nu}_{Ri}^{c}\nu_{Ri}.\label{eq:lag2}\end{align}
Thus, including the top mass $M_{t}$, there are seven unknown parameters 
\be M_{t}\h{2mm},\h{2mm}m_{\text{DM}}\h{2mm},\h{2mm}\kappa\h{2mm},\h{2mm}\lambda_{\text{DM}}\h{2mm},\h{2mm}y_{\nu}\h{2mm},\h{2mm}M_{R}\h{2mm},\h{2mm}Y_{R},\label{eq:para}\e
in this model. In the following discussion, to distinguish the initial values of these parameters at $\mu=M_{t}$ from their running couplings, we put the subscript 0 for their initial values, like $\kappa_{0}$ except for $M_{t}$. Because $S$ is the candidate of the DM, $m_{\text{DM}}$ and $\kappa$ must satisfy some relation such that they can explain the observed energy density \cite{Ade:2013zuv}
\be \Omega_{DM}h^{2}:=\frac{\rho_{\text{DM}}h^{2}}{\rho_{\text{tot}}}=0.1196\pm0.0031(\h{1mm}\text{68$\%$ CL}).\e
For $m_{\text{DM}}\gtrsim M_{h}$, this relation is approximately given by \cite{Cline:2013gha}
\be \log_{10}\kappa\simeq-3.63+1.04\log_{10}\left(\frac{m_{\text{DM}}}{\text{GeV}}\right).\label{eq:DMmass}\e  
Moreover, if we assume that the neutrino mass is $0.1$eV, $y_{\nu}$ and $M_{R}$ must satisfy 
\be -\frac{M_{R}}{2}\left(1-\sqrt{1+\frac{2y_{\nu}^{2}v_{h}^{2}}{M_{R}^{2}}}\right)\simeq\frac{y_{\nu}^{2}v_{h}^{2}}{2M_{R}}=0.1\text{eV},\label{eq:nmass}\e
where $v_{h}$ is the Higgs expectation value. This is the usual relation of the seesaw mechanism. In the following discussion, we choose $M_{R}=10^{13}$GeV, so $ y_{\nu}$ is fixed by Eq.(\ref{eq:nmass}). As a result, among the seven parameters, four of them remain as free parameters; they are
\be M_{t}\h{2mm},\h{2mm}\kappa\h{2mm},\h{2mm}\lambda_{\text{DM}}\h{2mm}\text{and}\h{2mm}Y_{R}.\e

To discuss how the effective couplings behave at the high energy scale, we must know the renormalization group equations (RGEs) of this model.  Their results are presented in \ref{app:beta}. Here, note that the contributions from the heavy right-handed neutrinos should be taken into account at the scale where $\mu\geq M_{R}$.
The 1-loop effective potential of the Higgs field is given by
\begin{align} V_{\text{1loop}}(\phi,\mu):=\begin{cases}V^{\text{SM}}_{\text{1loop}}(\phi)+\frac{M_{\text{DM}}(\phi)^{4}}{64\pi^{2}}\left[\log\left(\frac{M_{\text{DM}}(\phi)^{2}}{\mu^{2}}\right)-\frac{3}{2}\right]-6\cdot\frac{M^{-}_{\nu}(\phi)^{4}}{64\pi^{2}}\left[\log\left(\frac{M^{-}_{\nu}(\phi)^{2}}{\mu^{2}}\right)-\frac{3}{2}\right]\\
\h{10cm}(\text{for $\phi<M_{R}$}),\\
\\
V^{\text{SM}}_{\text{1loop}}(\phi)+\frac{M_{\text{DM}}(\phi)^{4}}{64\pi^{2}}\left[\log\left(\frac{M_{\text{DM}}(\phi)^{2}}{\mu^{2}}\right)-\frac{3}{2}\right]-6\cdot\frac{M^{-}_{\nu}(\phi)^{4}}{64\pi^{2}}\left[\log\left(\frac{M^{-}_{\nu}(\phi)^{2}}{\mu^{2}}\right)-\frac{3}{2}\right]\\
\\
\h{4cm}-6\cdot\frac{M^{+}_{\nu}(\phi)^{4}}{64\pi^{2}}\left[\log\left(\frac{M^{+}_{\nu}(\phi)^{2}}{\mu^{2}}\right)-\frac{3}{2}\right]\h{5mm}(\text{for $\phi>M_{R}$}),\end{cases}\end{align}
where
\be M_{\text{DM}}(\phi):=\sqrt{e^{2\Gamma(\phi)}\frac{\kappa\phi^{2}}{2}+m^{2}_{\text{DM}}}\h{2mm},\h{2mm} M_{\nu}^{\pm}(\phi):=\frac{M_{R}}{2}\left(1\pm\sqrt{1+\frac{2y_{\nu}^{2}e^{2\Gamma(\phi)}\phi^{2}}{M_{R}^{2}}}\right).\label{eq:effmass}\e

\begin{figure}
\begin{center}
\begin{tabular}{c}
\begin{minipage}{0.5\hsize}
\begin{center}
\includegraphics[width=8.5cm]{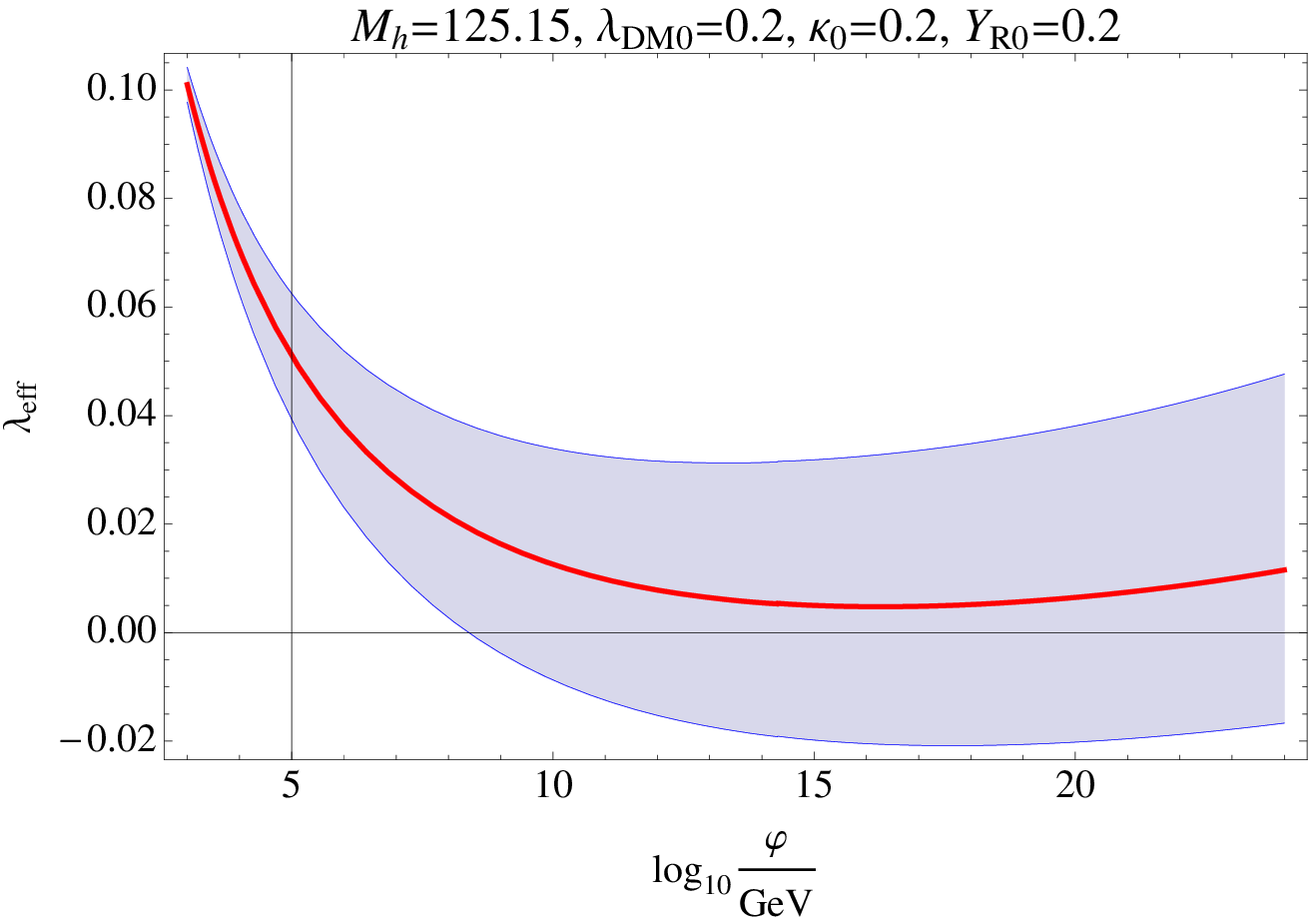}
\end{center}
\end{minipage}
\begin{minipage}{0.5\hsize}
\begin{center}
\includegraphics[width=8.5cm]{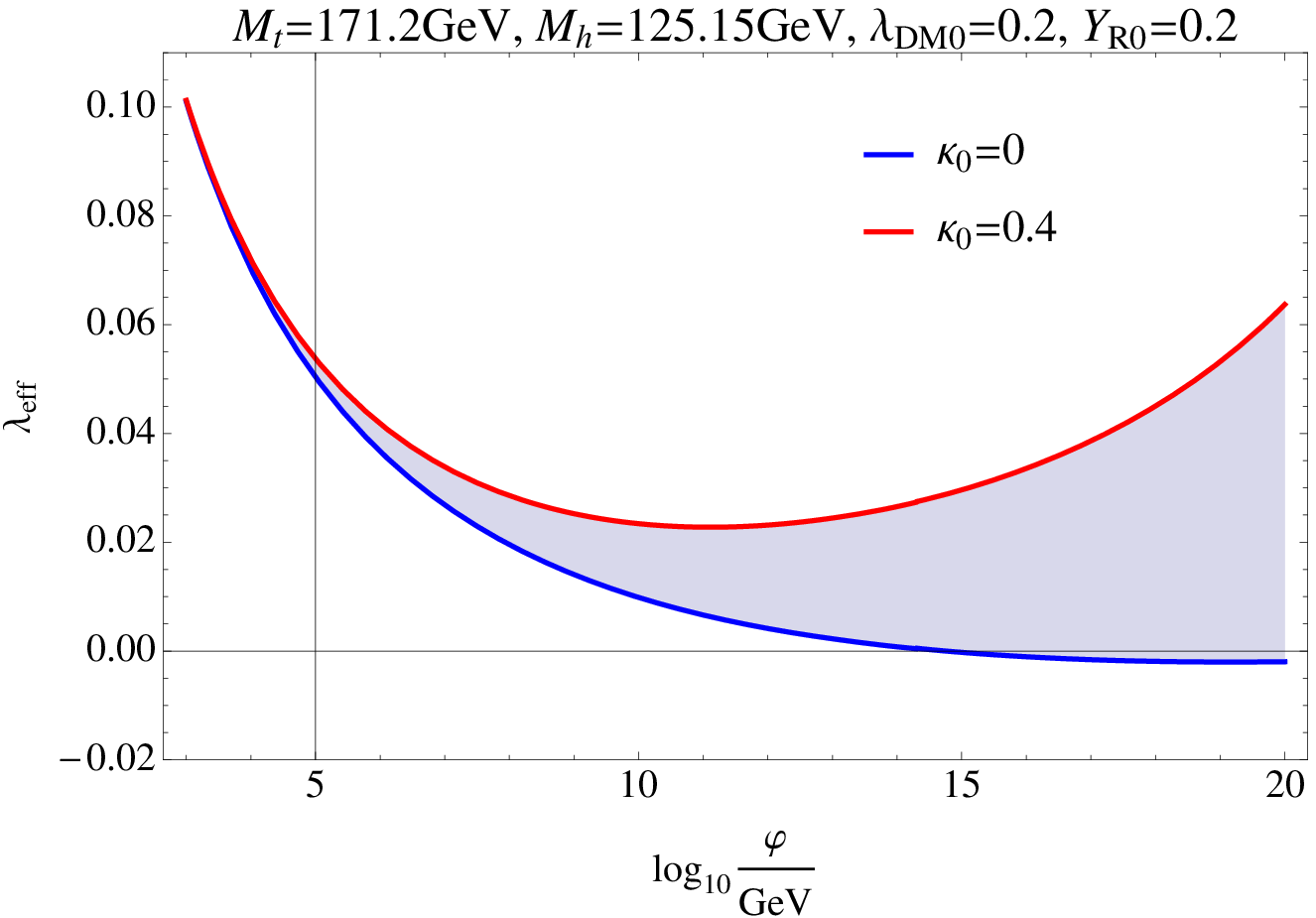}
\end{center}
\end{minipage}
\\
\\
\begin{minipage}{0.5\hsize}
\begin{center}
\includegraphics[width=8.5cm]{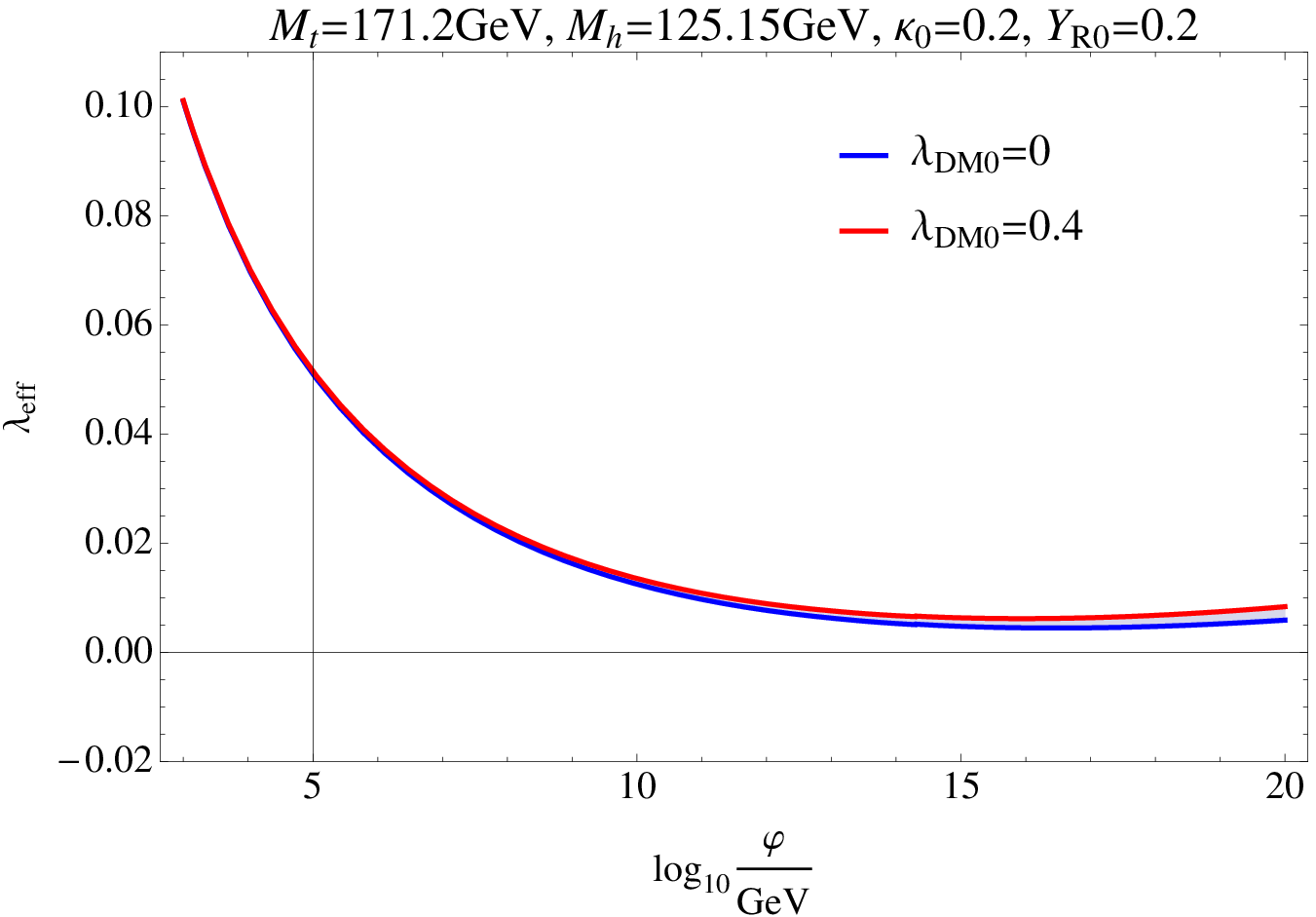}
\end{center}
\end{minipage}
\begin{minipage}{0.5\hsize}
\begin{center}
\includegraphics[width=8.5cm]{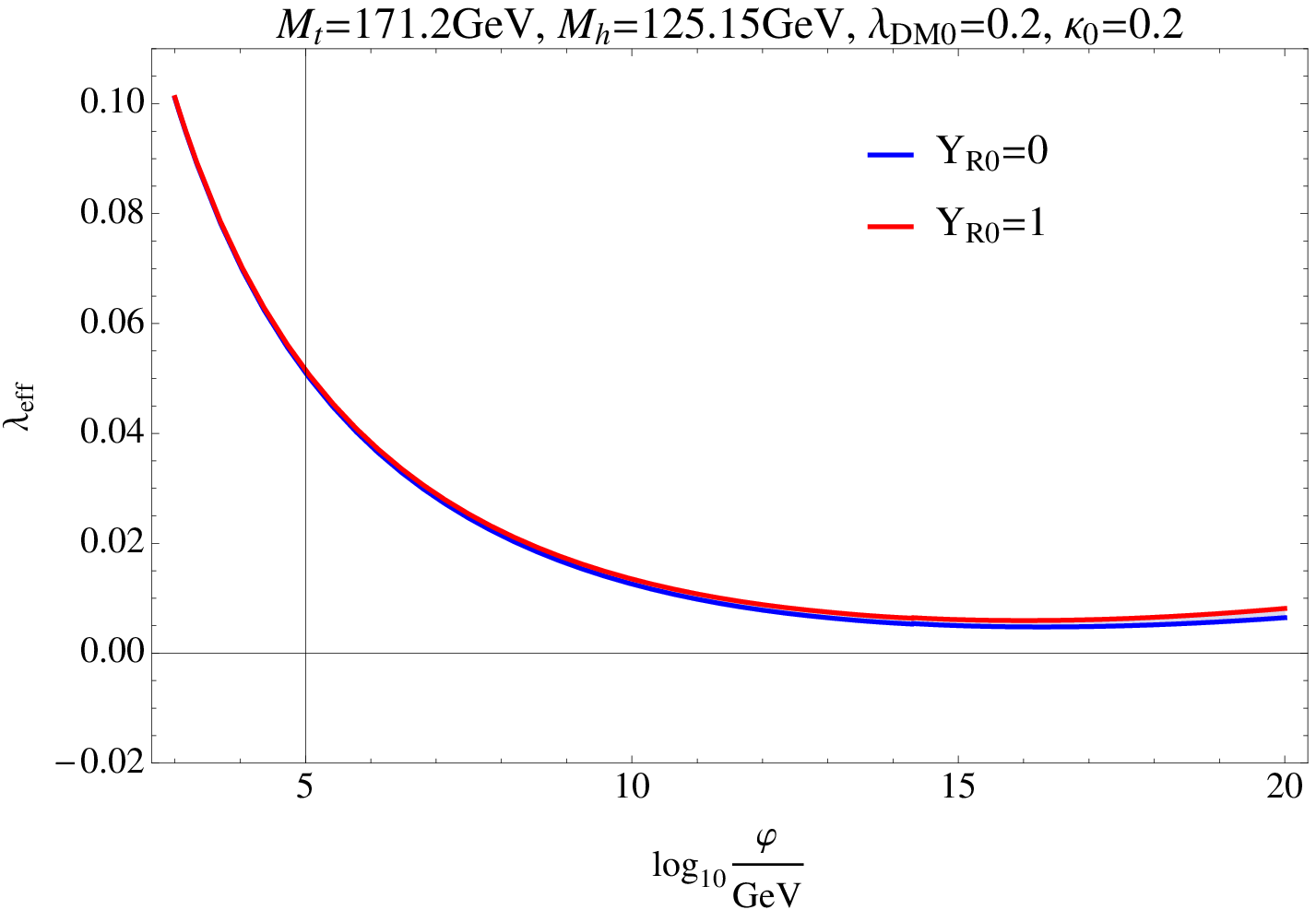}
\end{center}
\end{minipage}
\end{tabular}
\end{center}
\caption{The running effective Higgs self coupling $\lambda_{\text{eff}}$ as a function of $\phi$. The upper left (right) panel shows the $M_{t}\h{1mm}(\kappa_{0})$ dependence. For $M_{t}$, the blue band corresponds 95$\%$ CL deviation from $171.2$GeV. The lower left (right) panel shows the $\lambda_{\text{DM}0}$ $(Y_{R0})$ dependence.}
\label{fig:typical}
\end{figure}

\noindent In these expressions, we have put $S=0$ because 
we now focus on the MPP of the Higgs sector\footnote{Of course, we can consider the MPP of the DM sector. We study such situation in \ref{app:MPP}.}. Furthermore, we can neglect $m_{\text{DM}}$ in Eq.(\ref{eq:effmass}) because its effect is very small when $\phi\gg m_{\text{DM}}$. 
As well as Section\ref{sec:SM}, we put $\phi=\mu$, and define the effective Higgs self coupling $\lambda_{\text{eff}}$ as
\be \lambda_{\text{eff}}(\phi):=\frac{4}{\phi^{4}}V(\phi)=\frac{4}{\phi^{4}}\left(V_{\text{tree}}(\phi)+V_{\text{1loop}}(\phi)\right).\e
Fig.\ref{fig:typical} shows $\lambda_{\text{eff}}(\phi)$ for the various values of parameters. Here, the typical values are chosen to be
\be \lambda_{\text{DM}0}=0.2\h{2mm},\h{2mm}\kappa_{0}=0.2\h{2mm},\h{2mm}Y_{R0}=0.2.\e
One can see that $\lambda_{\text{eff}}$ depends mainly on $M_{t}$ and $\kappa_{0}$, and hardly on $\lambda_{\text{DM}0}$ and $Y_{R0}$. This is because $\lambda_{\text{DM}}$ does not appear in $\beta_{\lambda}$ and $Y_{R}$ appears at the two-loop level (see Eq.(\ref{eq:betalam}) in \ref{app:beta}). Therefore, by fixing $\lambda_{\text{DM}}$ and $Y_{R}$, we can relate $M_{t}$ and $\kappa_{0}$ from the MPP. 

By the same procedure of Section\ref{sec:SM}, we can calculate the scale $\Lambda_{\beta}$ where $\beta_{\lambda_{\text{eff}}}$ becomes zero, and obtain $\lambda_{\text{eff}}(\Lambda_{\beta})$ as a function of $M_{t}$ and $\kappa_{0}$. Fig.\ref{fig:region} shows the results. In the upper (lower) panels, $Y_{R0}$ is fixed to $0.2$ $(0.7)$. The difference between the left and right panels is whether the tree or one-loop level potential is used. 
The parameter region where $\lambda_{\text{eff}}(\Lambda_{\beta})<0$ and $\lambda_{\text{DM}}(\Lambda_{\beta})<0$ are filled respectively by blue and red. Both of them are excluded from the stability of the potentials. The MPP predicts that $M_{t}$ and $\kappa_{0}$ should exist on the green contour. One of the good points of this model is that the larger $M_{t}$ is allowed unlike the SM. This is consistent with the recent experimental value \cite{ATLAS:2014wva}
\be M_{t}=173.34\pm0.76\text{GeV},\label{eq:exp}\e 
which corresponds to the DM mass (see Eq.(\ref{eq:DMmass}))
\be 769\h{1mm}\text{GeV}\leq m_{\text{DM}}\leq 1053\h{1mm}\text{GeV}.\e
Two comments are needed. 
\begin{enumerate}\item The contours which represent $\Lambda_{\beta}=10^{16}$GeV, $10^{17}$GeV and $10^{18}$GeV are also shown in Fig.3 respectively by red, blue and orange. Thus, the larger $M_{t}$ (such as Eq.(\ref{eq:exp})) means that, in this model, the MPP of the Higgs potential occurs at the relatively low energy scale ($\lesssim 10^{16}$ GeV).\\
\item 
As is seen from the lower panels of Fig.\ref{fig:region}, we can also require $\lambda_{\text{DM}}(\Lambda_{\beta})=0$ in addition to $\lambda_{\text{eff}}(\lambda_{\beta})=0$. Because $\kappa_{0}$ and $Y_{R0}$ appear in the one-loop part of $\beta_{\lambda_{\text{DM}}}$, we can obtain a further relation between them by $\lambda_{\text{DM}}(\Lambda_{\beta})=0$. Although one might think that the remaining one parameter can be determined by $\beta_{\lambda_{\text{DM}}}(\Lambda_{\beta})=0$, we have checked that it is difficult to satisfy $\lambda_{\text{DM}}(\Lambda_{\beta})=\beta_{\lambda_{\text{DM}}}(\Lambda_{\beta})=0$ simultaneously. See \ref{app:MPP} for more details.\end{enumerate}

\begin{figure}
\begin{center}
\begin{tabular}{c}
\begin{minipage}{0.5\hsize}
\begin{center}
\includegraphics[width=8.2cm]{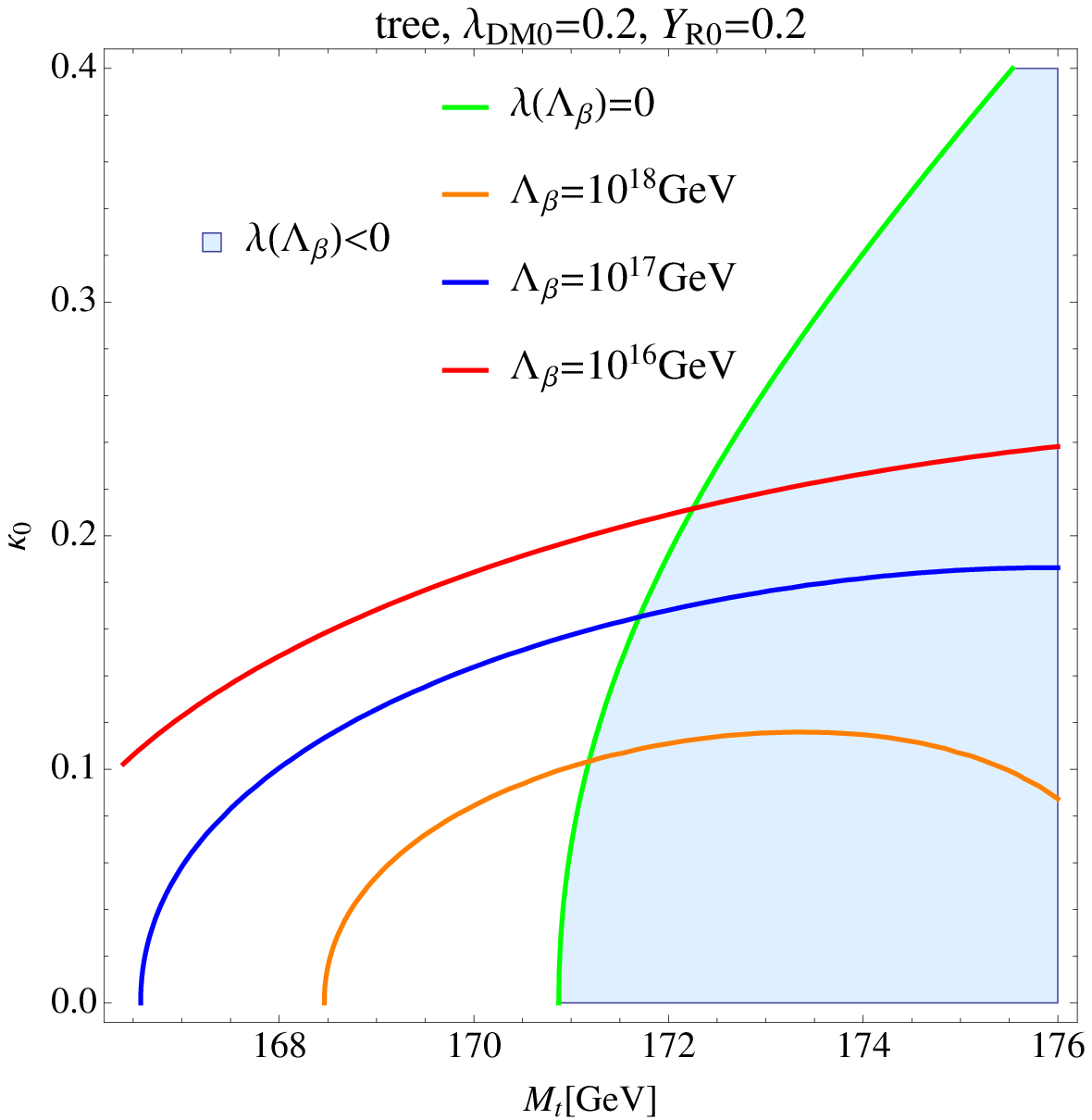}
\end{center}
\end{minipage}
\label{fig:region}
\begin{minipage}{0.5\hsize}
\begin{center}
\includegraphics[width=8.2cm]{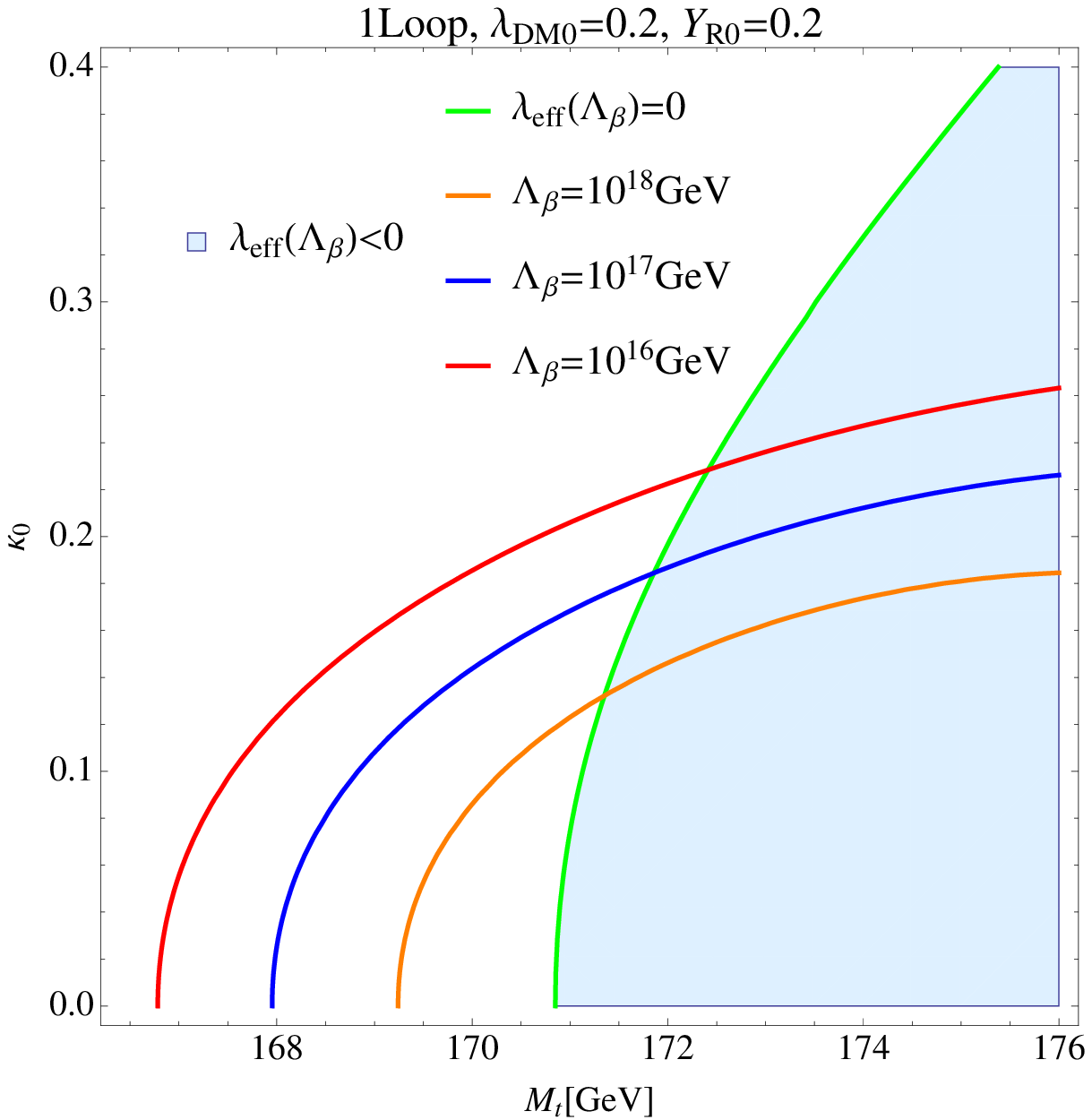}
\end{center}
\end{minipage}
\\
\\
\begin{minipage}{0.5\hsize}
\begin{center}
\includegraphics[width=8.2cm]{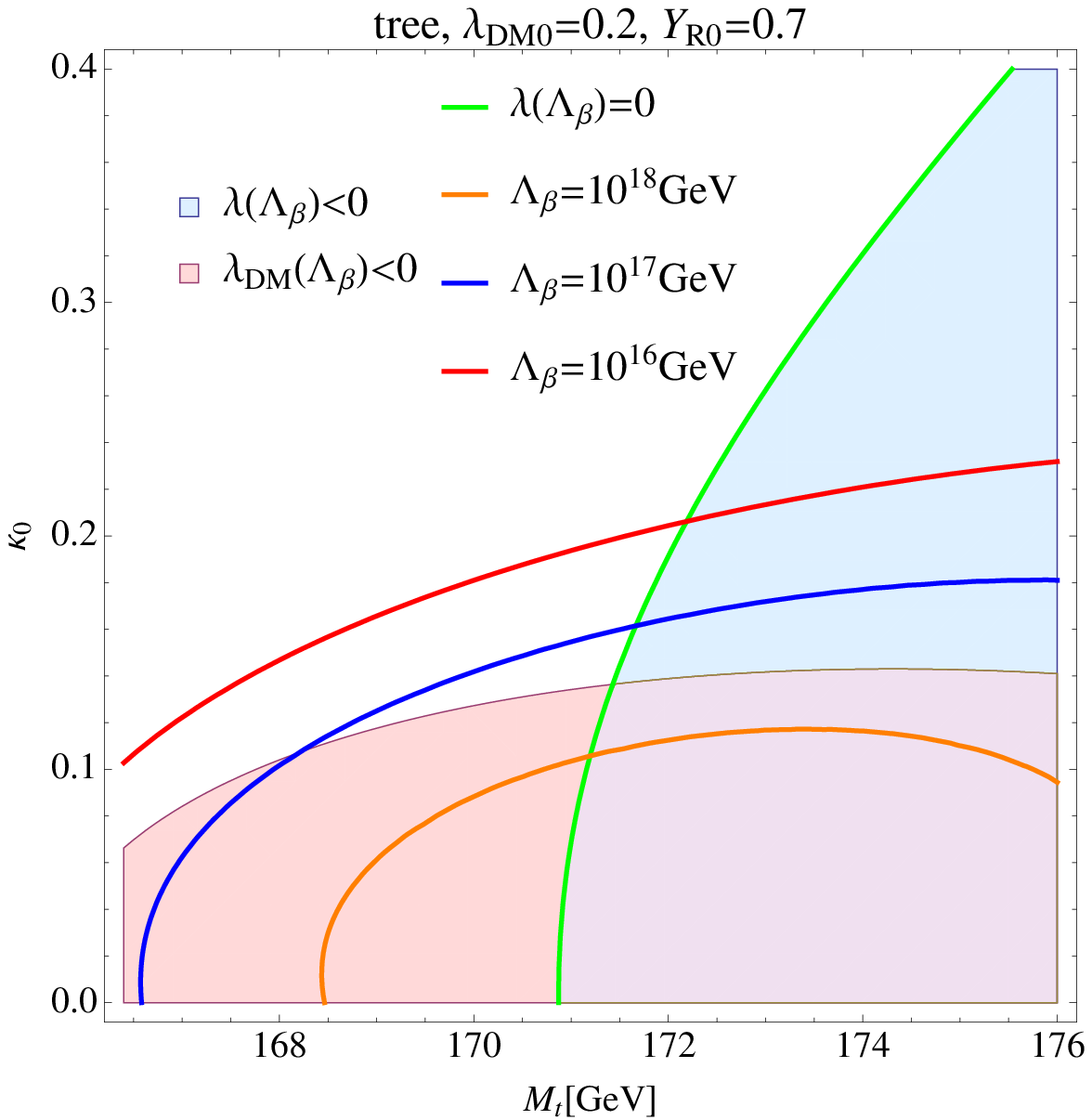}
\end{center}
\end{minipage}
\label{fig:region}
\begin{minipage}{0.5\hsize}
\begin{center}
\includegraphics[width=8.2cm]{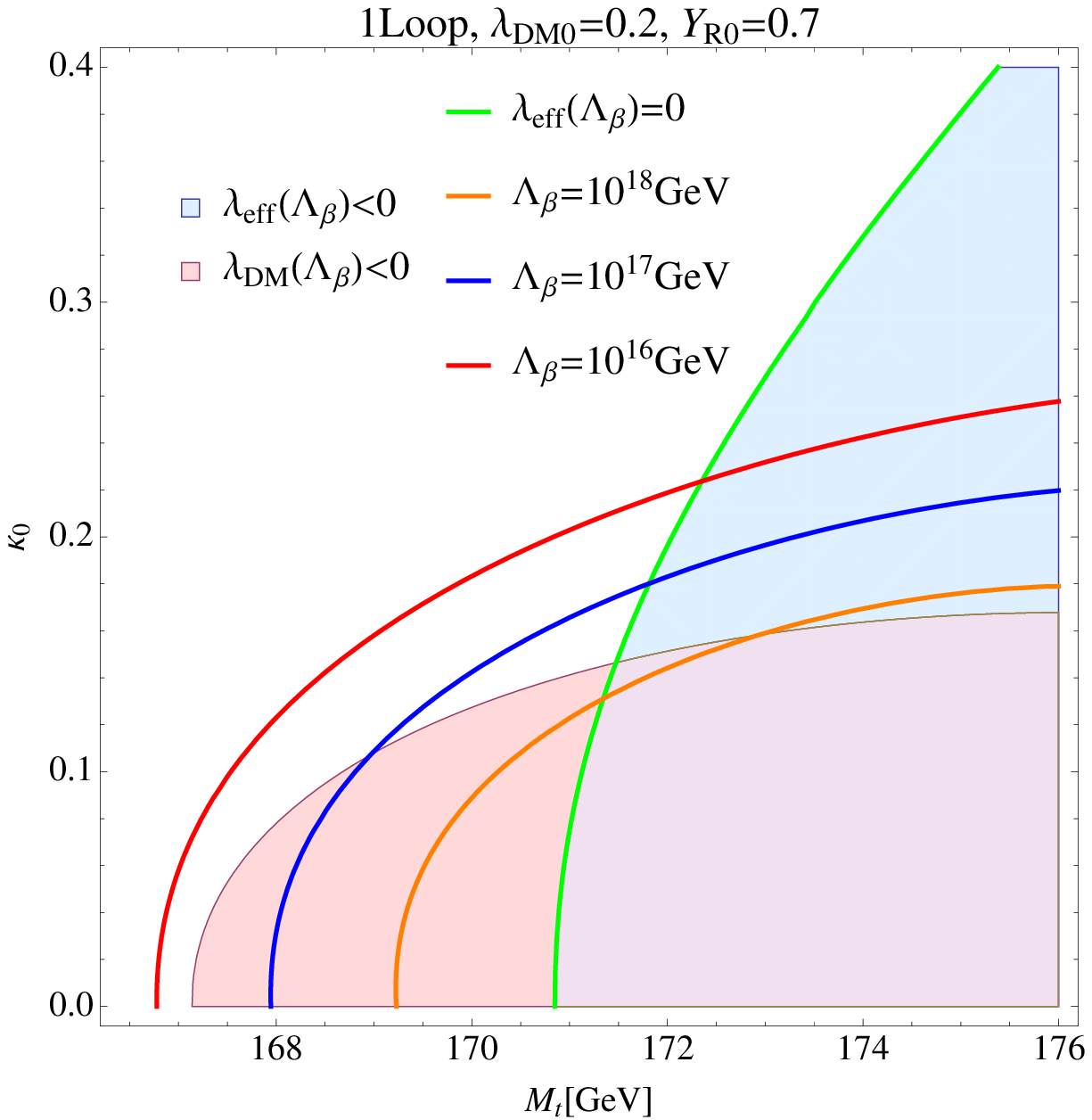}
\end{center}
\end{minipage}
\end{tabular}
\end{center}
\caption{The parameter dependences of $\lambda_{\text{eff}}(\Lambda_{\beta})$. Here, $\lambda_{\text{DM}0}$ is fixed to 0.2, and $Y_{R0}$ is fixed to $0.2(0.7)$ in the upper (lower) panels. The left (right) panels show the calculations by using the tree (one-loop) level potential. The green lines are the prediction by the MPP. The contours which represent $\Lambda_{\beta}=10^{16}$GeV, $10^{17}$GeV and $10^{18}$GeV are also shown respectively by red, blue and orange.} 
\end{figure}

\section{Summary}\label{sec:summary}
We have discussed the MPP of the SM with the scalar singlet DM and right-handed neutrinos. We have found that $\lambda_{\text{eff}}$ and $\beta_{\lambda_{\text{eff}}}$ can simultaneously become zero within the reasonable parameter region. The MPP predicts the strong relation between the portal coupling $\kappa$ and the top mass $M_{t}$. Unlike the pure SM, the larger $M_{t}$ is allowed in this model, which is favorable for the recent experimental values \cite{Moch:2014tta,ATLAS:2014wva}
\be M_{t}=173.34\pm0.76\text{GeV}.\e
Although we have found that the MPP can be satisfied for the Higgs potential, it is difficult to realize the exact flatness of the scalar potential at some high energy scale $\Lambda$;
\be \lambda(\Lambda)=\beta_{\lambda}(\Lambda)=\lambda_{\text{DM}}(\Lambda)=\beta_{\lambda_{\text{DM}}}(\Lambda)=\kappa(\Lambda)=\beta_{\kappa}(\Lambda)=0.\e
See \ref{app:MPP} for the details. It would be interesting to consider a generalization of this model in such a way that the MPP can be realized for the whole scalar fields.

\section*{Acknowledgement} 
We thank Hikaru Kawai, Yuta Hamada and Koji Tsumura for valuable discussions.

\appendix 
\def\thesection{Appendix \Alph{section}}
\section{Two Loop Renormalization Group Equations}\label{app:beta}
The two loop RGEs where the  Lagrangian is given by Eq.(\ref{eq:lag2}) are as follows:\footnote{The calculations in this appendix are based on \cite{Machacek:1983tz,Machacek:1983fi,Machacek:1984zw}, and our results are in agreement with the recent result \cite{Haba:2014zja} when there is only one right-handed neutrino and $Y_{R}=0$.}
\begin{align} \frac{dg_{Y}}{dt}&=\frac{1}{(4\pi)^{2}}\frac{41}{6}g_{Y}^{3}+\frac{g_{Y}^{3}}{(4\pi)^{4}}\left(\frac{199}{18}g_{Y}^{2}+\frac{9}{2}g_{2}^{2}+\frac{44}{3}g_{3}^{2}-\frac{17}{6}y_{t}^{2}-\frac{3}{2}y_{\nu}^{2}\right),
\\
\frac{dg_{2}}{dt}&=-\frac{1}{(4\pi)^{2}}\frac{19}{6}g_{2}^{3}+\frac{g_{2}^{3}}{(4\pi)^{4}}\left(\frac{3}{2}g_{Y}^{2}+\frac{35}{6}g_{2}^{2}+12g_{3}^{2}-\frac{3}{2}\left(y_{t}^{2}+y_{\nu}^{2}\right)\right),
\\
 \frac{dg_{3}}{dt}&=-\frac{7}{(4\pi)^{2}}g_{3}^{3}+\frac{g_{3}^{3}}{(4\pi)^{4}}\left(\frac{11}{6}g_{Y}^{2}+\frac{9}{2}g_{2}^{2}-26g_{3}^{2}-2y_{t}^{2}\right),
\\
\frac{dy_{t}}{dt}&=\frac{y_{t}}{(4\pi)^{2}}\left(\frac{9}{2}y_{t}^{2}+3y_{\nu}^{2}-\frac{17}{12}g_{Y}^{2}-\frac{9}{4}g_{2}^{2}-8g_{3}^{2}\right)\nonumber
\\
&+\frac{y_{t}}{(4\pi)^{4}}\Biggl\{-12y_{t}^{4}-\frac{27}{4}y_{\nu}^{4}-\frac{27}{4}y_{t}^{2}y_{\nu}^{2}-\frac{9}{8}Y_{R}^{2}y_{\nu}^{2}+6\lambda^{2}+\frac{1}{4}\kappa^{2}-12\lambda y_{t}^{2}+g_{Y}^{2}\left(\frac{131}{16}y_{t}^{2}+\frac{15}{8}y_{\nu}^{2}\right)\nonumber
\\
&\qquad+g_{2}^{2}\left(\frac{225}{16}y_{t}^{2}+\frac{45}{8}y_{\nu}^{2}\right)+36g_{3}^{2}y_{t}^{2}+\frac{1187}{216}g_{Y}^{4}-\frac{23}{4}g_{2}^{4}-108g_{3}^{4}-\frac{3}{4}g_{Y}^{2}g_{2}^{2}+9g_{2}^{2}g_{3}^{2}+\frac{19}{9}g_{3}^{2}g_{Y}^{2}\Biggl\},
\\
\frac{dy_{\nu}}{dt}&=\frac{y_{\nu}}{(4\pi)^{2}}\left(\frac{9}{2}y_{\nu}^{2}+3y_{t}^{2}+\frac{1}{4}Y_{R}^{2}-\frac{3}{4}g_{Y}^{2}-\frac{9}{4}g_{2}^{2}\right)\nonumber
\\
&+\frac{y_{\nu}}{(4\pi)^{4}}\Biggl\{-12y_{\nu}^{4}-\frac{27}{4}y_{t}^{4}-\frac{19}{32}Y_{R}^{4}-y_{\nu}^{2}\left(\frac{27}{4}y_{t}^{2}+\frac{21}{16}Y_{R}^{2}\right)+6\lambda^{2}+\frac{1}{4}\kappa^{2}-12\lambda y_{\nu}^{2}-\kappa Y_{R}^{2}\nonumber
\\
&\qquad+g_{Y}^{2}\left(\frac{123}{16}y_{\nu}^{2}+\frac{85}{24}y_{t}^{2}+\frac{9}{16}Y_{R}^{2}\right)+g_{2}^{2}\left(\frac{225}{16}y_{\nu}^{2}+\frac{45}{8}y_{t}^{2}+\frac{27}{16}Y_{R}^{2}\right)+20g_{3}^{2}y_{t}^{2}\nonumber
\\
&\qquad\h{5mm}+\frac{35}{24}g_{Y}^{4}-\frac{23}{4}g_{2}^{4}-\frac{9}{4}g_{Y}^{2}g_{2}^{2}\Biggl\},
\\
\frac{dY_{R}}{dt}&=\frac{Y_{R}}{(4\pi)^{2}}\left(3Y_{R}^{2}+2y_{\nu}^{2}\right)+\frac{Y_{R}}{(4\pi)^{4}}\Biggl\{-\frac{81}{16}Y_{R}^{4}-\frac{27}{4}Y_{R}^{2}y_{\nu}^{2}-9y_{t}^{2}y_{\nu}^{2}-\frac{27}{2}y_{\nu}^{4}+\frac{1}{12}\lambda_{\text{DM}}^{2}+\kappa^{2}\nonumber
\\
&\h{6cm}-\lambda_{\text{DM}}Y_{R}^{2}-8\kappa y_{\nu}^{2}-\frac{1}{4}g_{Y}^{2}y_{\nu}^{2}-\frac{3}{4}g_{2}^{2}y_{\nu}^{2}\Biggl\},
\end{align}
\newpage
\begin{align}\frac{d\lambda}{dt}&=\frac{1}{(4\pi)^{2}}\left(\lambda \left(24 \lambda-9 g_{2}^2-3
   g_{Y}^2 +12 y_{\nu}^2+12
   y_{t}^2\right)+\frac{3}{4} g_{2}^2
   g_{Y}^2+\frac{9 g_{2}^4}{8}+\frac{3
   g_{Y}^4}{8}+\frac{\kappa ^2}{2}-6y_{\nu
   }^4-6 y_{t}^4\right)\nonumber
\\
&+\frac{1}{(4\pi)^{4}} \Biggl\{-2\kappa^{3}-5\kappa^{2}\lambda-312\lambda^{3}+36\lambda^{2}\left(g_{Y}^{2}+3g_{2}^{2}\right)+\lambda\left(\frac{629}{24}g_{Y}^{4}+\frac{39}{4}g_{2}^{2}g_{Y}^{2}-\frac{73}{8}g_{2}^{4}\right)\nonumber
\\
&\quad+\frac{305}{16}g_{2}^{6}-\frac{289}{48}g_{Y}^{2}g_{2}^{4}-\frac{559}{48}g_{Y}^{4}g_{2} ^{2}+\frac{379}{48}g_{Y}^{6}-32g_{3}^{2}y_{t}^{4}-\frac{8}{3}g_{Y}^{2}y_{t}^{4}-\frac{9}{4}g_{2}^{4}\left(y_{t}^{2}+y_{\nu}^{2}\right)\nonumber
\\
&\quad+\lambda y_{t}^{2}\left(\frac{85}{6}g_{Y}^{2}+\frac{45}{2}g_{2}^{2}+80g_{3}^{2}\right)+\lambda y_{\nu}^{2}\left(\frac{15}{2}g_{Y}^{2}+\frac{45}{2}g_{2}^{2}\right)+g_{Y}^{2}y_{t}^{2}\left(-\frac{19}{4}g_{Y}^{2}+\frac{21}{2}g_{2}^{2}\right)\nonumber
\\
&\quad-g_{Y}^{2}y_{\nu}^{2}\left(\frac{3}{4}g_{Y}^{2}+\frac{3}{2}g_{2}^{2}\right)-144\lambda^{2}\left(y_{t}^{2}+y_{\nu}^{2}\right)-3\lambda\left(y_{t}^{4}+y_{\nu}^{4}+\frac{3}{2}Y_{R}^{2}y_{\nu}^{2}\right)\nonumber
\\
&\h{7cm}+30\left(y_{t}^{6}+y_{\nu}^{6}+\frac{1}{5}Y_{R}^{2}y_{\nu}^{4}\right)-\frac{3}{2}Y_{R}^{2}\kappa^{2}\Biggl\},\label{eq:betalam}
\\
\frac{d\lambda_{\text{DM}}}{dt}&=\frac{1}{(4\pi)^{2}}\left(3\lambda_{\text{DM}}^{2}+12\kappa^{2}+6\lambda_{\text{DM}}Y_{R}^{2}-18Y_{R}^{4}\right)\nonumber
\\
&+\frac{1}{(4\pi)^{4}}\Biggl\{-\frac{17}{3}\lambda_{\text{DM}}^{3}-20\kappa^{2}\lambda_{\text{DM}}-48\kappa^{3}-72\left(y_{t}^{2}+y_{\nu}^{2}\right)\kappa^{2}+24\left(g_{Y}^{2}+3g_{2}^{2}\right)\kappa^{2}
\nonumber
\\
&\qquad+72Y_{R}^{4}\left(Y_{R}^{2}+y_{\nu}^{2}\right)+\lambda_{\text{DM}}Y_{R}^{2}\left(\frac{21}{2}Y_{R}^{2}-18y_{\nu}^{2}\right)-9Y_{R}^{2}\lambda_{\text{DM}}^{2}\Biggl\},
\\
\frac{d\kappa}{dt}&=\frac{1}{(4\pi)^{2}}\left(4\kappa^{2}+12\kappa\lambda+\kappa\lambda_{\text{DM}}+3\kappa\left(2y_{t}^{2}+2y_{\nu}^{2}+Y_{R}^{2}\right)-\frac{3}{2}\kappa\left(g_{Y}^{2}+3g_{2}^{2}\right)-12Y_{R}^{2}y_{\nu}^{2}\right)\nonumber
\\
&+\frac{\kappa}{(4\pi)^{4}}\Biggl\{-\frac{21}{2}\kappa^{2}-72\kappa\lambda-60\lambda^{2}-6\kappa\lambda_{\text{DM}}-\frac{5}{6}\lambda_{\text{DM}}^{2}-\left(y_{t}^{2}+y_{\nu}^{2}\right)\left(12\kappa+72\lambda\right)-3Y_{R}^{2}\left(2\kappa+\lambda_{\text{DM}}\right)
\nonumber
\\
&\quad-\frac{27}{2}y_{t}^{4}-\frac{27}{2}y_{\nu}^{4}-\frac{3}{4}Y_{R}^{4}+\frac{51}{4}Y_{R}^{2}y_{\nu}^{2}+g_{Y}^{2}\left(\kappa+24\lambda\right)+3g_{2}^{2}\left(\kappa+24\lambda\right)+y_{t}^{2}\left(\frac{85}{12}g_{Y}^{2}+\frac{45}{4}g_{2}^{2}+40g_{3}^{2}\right)\nonumber
\\
&\quad+y_{\nu}^{2}\left(\frac{15}{4}g_{Y}^{2}+\frac{45}{4}g_{2}^{2}\right)+\frac{557}{48}g_{Y}^{4}-\frac{145}{16}g_{2}^{4}+\frac{15}{8}g_{Y}^{2}g_{2}^{2}\Biggl\}+\frac{Y_{R}^{2}y_{\nu}^{2}}{(4\pi)^{4}}\Biggl\{\frac{3}{2}\left(g_{Y}^{2}+3g_{2}^{2}\right)+27Y_{R}^{2}+66y_{\nu}^{2}\Biggl\},
\\
\frac{d\Gamma}{dt}&=\frac{1}{(4\pi)^{2}}\left(\frac{9}{4}g_{2}^{2}+\frac{3}{4}g_{Y}^{2}-3y_{t}^{2}-3y_{\nu}^{2}\right).
\end{align}

\newpage
\section{Is Exact Flat Potential Possible?}\label{app:MPP}
\begin{figure}
\begin{minipage}{0.5\hsize}
\begin{center}
\includegraphics[width=7cm]{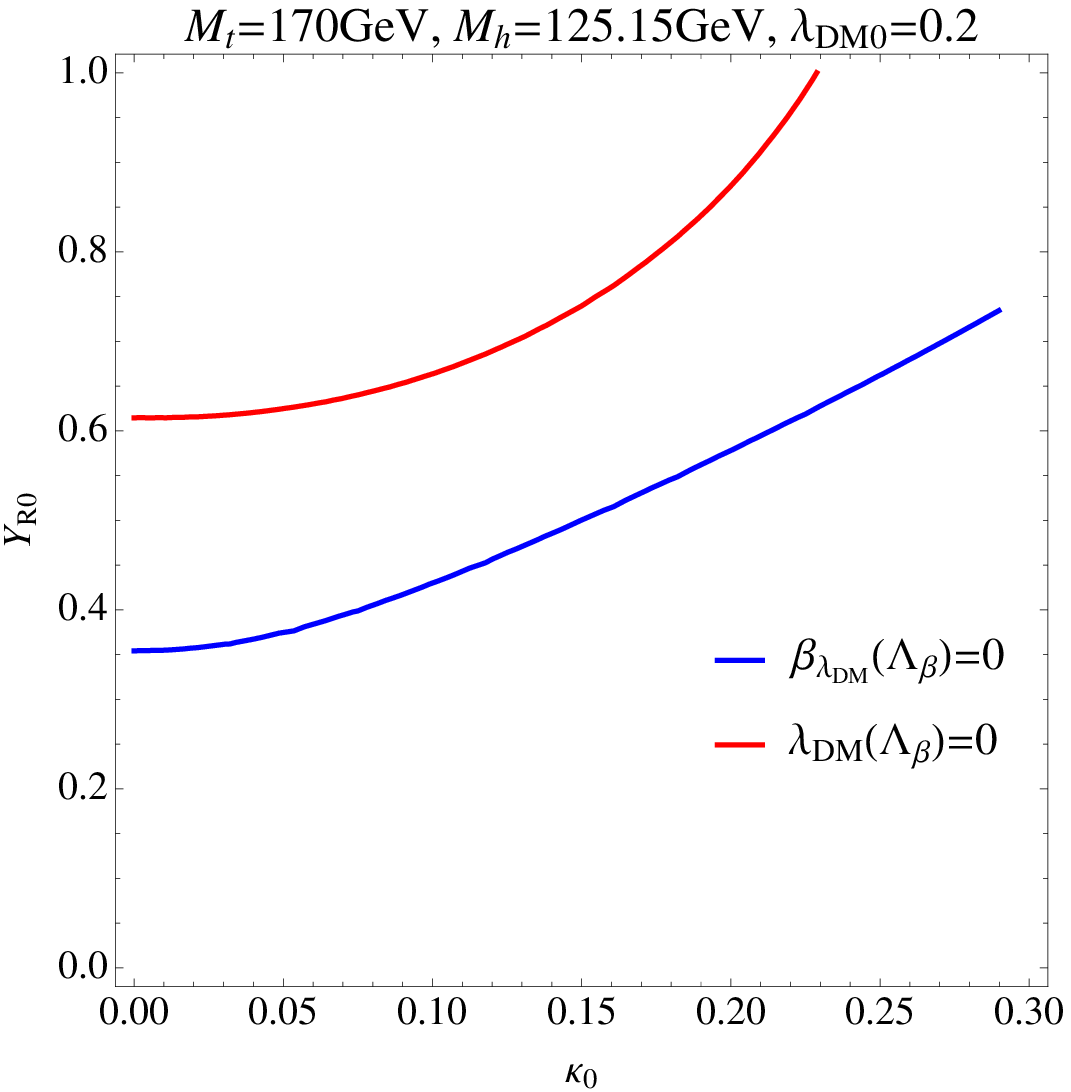}
\end{center}
\end{minipage}
\label{fig:contour}
\begin{minipage}{0.5\hsize}
\begin{center}
\includegraphics[width=7cm]{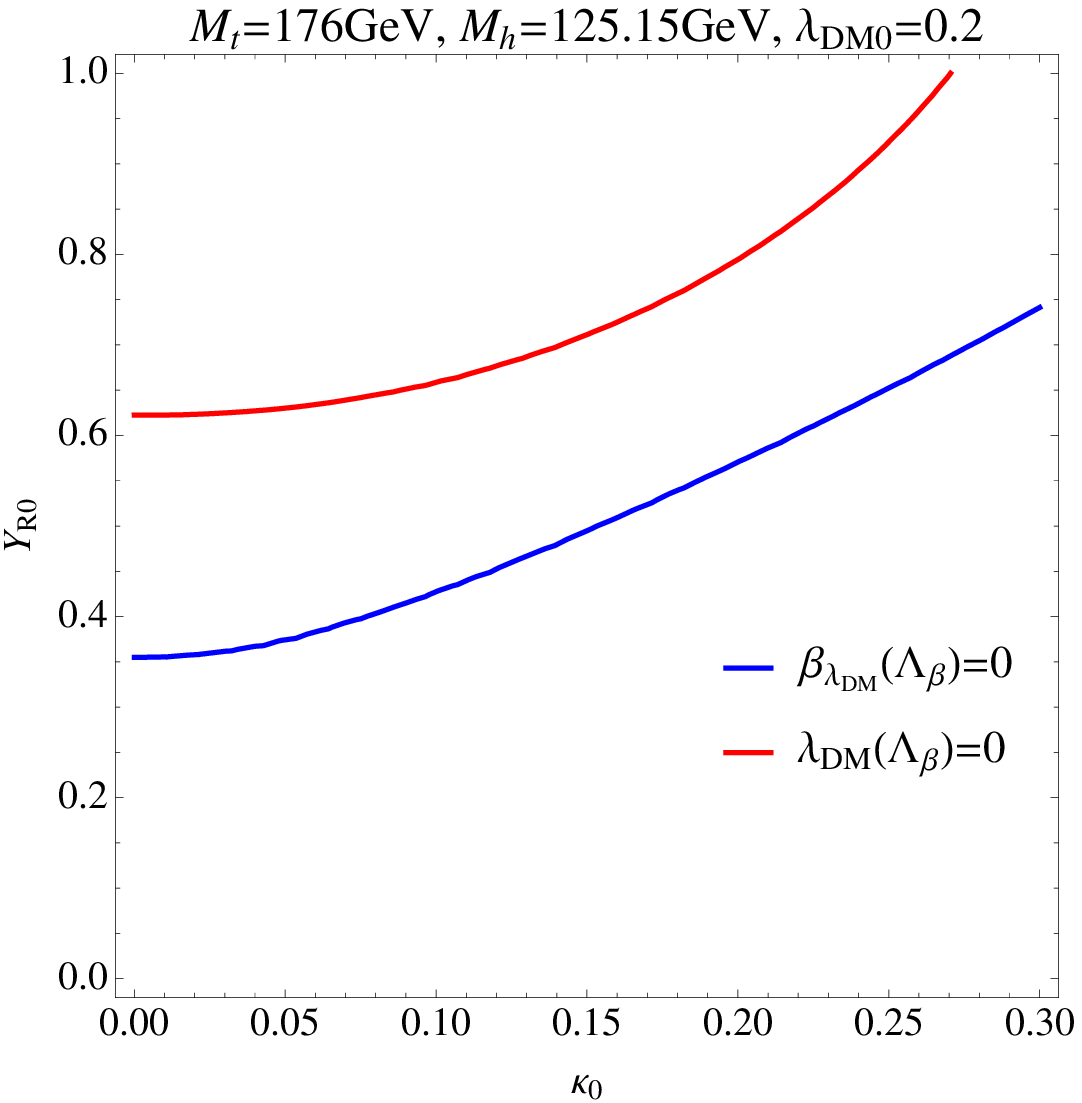}
\end{center}
\end{minipage}
\caption{The blue (red) lines show the contours where $\beta_{\lambda_{\text{DM}}}(\lambda_{\text{DM}})(\Lambda_{\beta})=0$. The left (right) panel is the $M_{t}=170(176)$GeV case. Here, note that if $\kappa_{0}\gtrsim0.3$, $\Lambda_{\beta}$ becomes less than $M_{R}=10^{13}$ GeV, and there is no solution of $\beta_{\lambda_{\text{DM}}}(\Lambda_{\beta})=0$ because the one-loop part of $\beta_{\lambda_{\text{DM}}}$ is always positive when $\mu\leq M_{R}$. } 
\end{figure}

One of the question is whether the MPP can be realized exactly. Namely,
\be \lambda(\Lambda_{\beta})=\beta_{\lambda}(\Lambda_{\beta})=\lambda_{\text{DM}}(\Lambda_{\beta})=\beta_{\lambda_{\text{DM}}}(\Lambda_{\beta})=\kappa(\Lambda_{\beta})=\beta_{\kappa}(\Lambda_{\beta})=0\label{eq:exaMPP}\e
is possible or not. Here, for simplicity, we also define $\Lambda_{\beta}$ as the scale where $\beta_{\lambda}$ becomes zero. To discuss this possibility, it is qualitatively enough to consider the one-loop RGEs. One can easily understand it is impossible to realize Eq.(\ref{eq:exaMPP}) as follows; even if $\lambda(\Lambda_{\beta})$, $\beta_{\lambda}(\Lambda_{\beta})$, $\lambda_{\text{DM}}(\Lambda_{\beta})$ and $\beta_{\lambda_{\text{DM}}}(\Lambda_{\beta})$ become simultaneously zero, we can not make $\kappa(\Lambda_{\beta})$ zero because the one-loop part of $\beta_{\lambda_{\text{DM}}}$ at $\Lambda_{\beta}$ becomes 
\be \beta_{\lambda_{\text{DM}}}|_{\Lambda_{\beta}}=\frac{1}{(4\pi)^{2}}\left(12\kappa^{2}-18Y_{R}^{4}\right),
\e
and we need $\kappa(\Lambda_{\beta})\neq0$ to satisfy $\beta_{\lambda_{\text{DM}}}(\Lambda_{\beta})=0$ \footnote{Typically, the non-zero positive  $Y_{R0}$ is needed to make $\beta_{\lambda_{\text{DM}}}$ negative. As a result, $Y_{R}(\Lambda_{\beta})$ is non-zero because the one-loop part of $\beta_{Y_{R}}$ is always positive.}.  Furthermore, it is also difficult even to satisfy $\lambda_{\text{DM}}(\Lambda_{\beta})=\beta_{\lambda_{\text{DM}}}(\Lambda_{\beta})=0$ simultaneously. See Fig.4. This shows the contours such that $\lambda_{\text{DM}}(\Lambda_{\beta})$ and $\beta_{\lambda_{\text{DM}}}(\Lambda_{\beta})$ become zero respectively. Here, we have used the two-loop RGEs. One can see that two contours do not intersect.


\begin{thebibliography}{unsrt}

\bibitem{Aad:2012tfa} 
  G.~Aad {\it et al.}  [ATLAS Collaboration],
  ``Observation of a new particle in the search for the Standard Model Higgs boson with the ATLAS detector at the LHC,''
  Phys.\ Lett.\ B {\bf 716}, 1 (2012)
  [arXiv:1207.7214 [hep-ex]].

\bibitem{Chatrchyan:2012ufa} 
  S.~Chatrchyan {\it et al.}  [CMS Collaboration],
  ``Observation of a new boson at a mass of 125 GeV with the CMS experiment at the LHC,''
  Phys.\ Lett.\ B {\bf 716}, 30 (2012)
  [arXiv:1207.7235 [hep-ex]].

\bibitem{Froggatt:1995rt} 
  C.~D.~Froggatt and H.~B.~Nielsen,
  ``Standard model criticality prediction: Top mass 173 +- 5-GeV and Higgs mass 135 +- 9-GeV,''
  Phys.\ Lett.\ B {\bf 368}, 96 (1996)
  [hep-ph/9511371].

\bibitem{Froggatt:2001pa} 
  C.~D.~Froggatt, H.~B.~Nielsen and Y.~Takanishi,
  ``Standard model Higgs boson mass from borderline metastability of the vacuum,''
  Phys.\ Rev.\ D {\bf 64}, 113014 (2001)
  [hep-ph/0104161].

\bibitem{Nielsen:2012pu} 
  H.~B.~Nielsen,
  ``PREdicted the Higgs Mass,''
  arXiv:1212.5716 [hep-ph].

\bibitem{Kawana:2015tka} 
  K.~Kawana,
  ``Criticality and Inflation of the Gauged B-L Model,''
  arXiv:1501.04482 [hep-ph].

\bibitem{Shaposhnikov:2009pv} 
  M.~Shaposhnikov and C.~Wetterich,
  ``Asymptotic safety of gravity and the Higgs boson mass,''
  Phys.\ Lett.\ B {\bf 683}, 196 (2010)
  [arXiv:0912.0208 [hep-th]].

\bibitem{Meissner:2007xv} 
  K.~A.~Meissner and H.~Nicolai,
  ``Effective action, conformal anomaly and the issue of quadratic divergences,''
  Phys.\ Lett.\ B {\bf 660}, 260 (2008)
  [arXiv:0710.2840 [hep-th]].

\bibitem{Khoze:2014xha} 
  V.~V.~Khoze, C.~McCabe and G.~Ro,
  ``Higgs vacuum stability from the dark matter portal,''
  JHEP {\bf 1408}, 026 (2014)
  [arXiv:1403.4953 [hep-ph], arXiv:1403.4953].

\bibitem{Kawai:2011qb} 
  H.~Kawai and T.~Okada,
  ``Solving the Naturalness Problem by Baby Universes in the Lorentzian Multiverse,''
  Prog.\ Theor.\ Phys.\  {\bf 127}, 689 (2012)
  [arXiv:1110.2303 [hep-th]].

\bibitem{Kawai:2013wwa} 
  H.~Kawai,
  ``Low energy effective action of quantum gravity and the naturalness problem,''
  Int.\ J.\ Mod.\ Phys.\ A {\bf 28}, 1340001 (2013).

\bibitem{Hamada:2014ofa} 
  Y.~Hamada, H.~Kawai and K.~Kawana,
  ``Evidence of the Big Fix,''
  Int.\ J.\ Mod.\ Phys.\ A {\bf 29}, no. 17, 1450099 (2014)
  [arXiv:1405.1310 [hep-ph]].

\bibitem{Hamada:2014xra} 
  Y.~Hamada, H.~Kawai and K.~Kawana,
  ``Weak Scale From the Maximum Entropy Principle,''
  arXiv:1409.6508 [hep-ph].

\bibitem{Hamada:2013mya} 
  Y.~Hamada, H.~Kawai and K.~y.~Oda,
  ``Minimal Higgs inflation,''
  PTEP {\bf 2014}, 023B02 (2014)
  [arXiv:1308.6651 [hep-ph]].

\bibitem{Hamada:2014iga} 
  Y.~Hamada, H.~Kawai, K.~y.~Oda and S.~C.~Park,
  ``Higgs inflation still alive,''
  Phys.\ Rev.\ Lett.\  {\bf 112}, 241301 (2014)
  [arXiv:1403.5043 [hep-ph]].

\bibitem{Hamada:2014wna} 
  Y.~Hamada, H.~Kawai, K.~y.~Oda and S.~C.~Park,
  ``Higgs inflation from Standard Model criticality,''
  arXiv:1408.4864 [hep-ph].

\bibitem{Hamada:2014raa} 
  Y.~Hamada, K.~y.~Oda and F.~Takahashi,
  ``Topological Higgs inflation: The origin of the Standard Model criticality,''
  arXiv:1408.5556 [hep-ph].

\bibitem{Kawamura:2013kua} 
  Y.~Kawamura,
  ``Naturalness, Conformal Symmetry and Duality,''
  PTEP {\bf 2013}, no. 11, 113B04 (2013)
  [arXiv:1308.5069 [hep-ph]].

\bibitem{Meissner:2006zh} 
  K.~A.~Meissner and H.~Nicolai,
  ``Conformal Symmetry and the Standard Model,''
  Phys.\ Lett.\ B {\bf 648}, 312 (2007)
  [hep-th/0612165].

\bibitem{Haba:2014sia} 
  N.~Haba, H.~Ishida, K.~Kaneta and R.~Takahashi,
  ``Vanishing Higgs potential at the Planck scale in a singlet extension of the standard model,''
  Phys.\ Rev.\ D {\bf 90}, 036006 (2014)
  [arXiv:1406.0158 [hep-ph]].

\bibitem{Iso:2012jn} 
  S.~Iso and Y.~Orikasa,
  ``TeV Scale B-L model with a flat Higgs potential at the Planck scale - in view of the hierarchy problem -,''
  PTEP {\bf 2013}, 023B08 (2013)
  [arXiv:1210.2848 [hep-ph]].

\bibitem{Hamada:2015ria} 
  Y.~Hamada, H.~Kawai and K.~y.~Oda,
  ``Eternal Higgs inflation and cosmological constant problem,''
  arXiv:1501.04455 [hep-ph].

\bibitem{Hamada:2014xka} 
  Y.~Hamada, H.~Kawai and K.~y.~Oda,
  ``Predictions on mass of Higgs portal scalar dark matter from Higgs inflation and flat potential,''
  JHEP {\bf 1407}, 026 (2014)
  [arXiv:1404.6141 [hep-ph], arXiv:1404.6141].

\bibitem{Haba:2014zda} 
  N.~Haba and R.~Takahashi,
  ``Higgs inflation with singlet scalar dark matter and right-handed neutrino in light of BICEP2,''
  Phys.\ Rev.\ D {\bf 89}, 115009 (2014)
  [arXiv:1404.4737 [hep-ph]].

\bibitem{Haba:2014zja} 
  N.~Haba, H.~Ishida and R.~Takahashi,
  ``Higgs inflation and Higgs portal dark matter with right-handed neutrinos,''
  arXiv:1405.5738 [hep-ph].

\bibitem{Buttazzo:2013uya} 
  D.~Buttazzo, G.~Degrassi, P.~P.~Giardino, G.~F.~Giudice, F.~Sala, A.~Salvio and A.~Strumia,
  ``Investigating the near-criticality of the Higgs boson,''
  JHEP {\bf 1312}, 089 (2013)
  [arXiv:1307.3536 [hep-ph]].


\bibitem{Moch:2014tta} 
  S.~Moch, S.~Weinzierl, S.~Alekhin, J.~Blumlein, L.~de la Cruz, S.~Dittmaier, M.~Dowling and J.~Erler {\it et al.},
  ``High precision fundamental constants at the TeV scale,''
  arXiv:1405.4781 [hep-ph].

\bibitem{ATLAS:2014wva} 
  [ATLAS and CDF and CMS and D0 Collaborations],
  ``First combination of Tevatron and LHC measurements of the top-quark mass,''
  arXiv:1403.4427 [hep-ex].

\bibitem{Ade:2013zuv} 
  P.~A.~R.~Ade {\it et al.}  [Planck Collaboration],
  ``Planck 2013 results. XVI. Cosmological parameters,''
  Astron.\ Astrophys.\  (2014)
  [arXiv:1303.5076 [astro-ph.CO]].

\bibitem{Cline:2013gha} 
  J.~M.~Cline, K.~Kainulainen, P.~Scott and C.~Weniger,
  ``Update on scalar singlet dark matter,''
  Phys.\ Rev.\ D {\bf 88}, 055025 (2013)
  [arXiv:1306.4710 [hep-ph]].

\bibitem{Hamada:2012bp} 
  Y.~Hamada, H.~Kawai and K.~y.~Oda,
  ``Bare Higgs mass at Planck scale,''
  Phys.\ Rev.\ D {\bf 87}, no. 5, 053009 (2013)
  [arXiv:1210.2538 [hep-ph]].

\bibitem{Machacek:1983tz} 
  M.~E.~Machacek and M.~T.~Vaughn,
  ``Two Loop Renormalization Group Equations in a General Quantum Field Theory. 1. Wave Function Renormalization,''
  Nucl.\ Phys.\ B {\bf 222}, 83 (1983).

\bibitem{Machacek:1983fi} 
  M.~E.~Machacek and M.~T.~Vaughn,
  ``Two Loop Renormalization Group Equations in a General Quantum Field Theory. 2. Yukawa Couplings,''
  Nucl.\ Phys.\ B {\bf 236}, 221 (1984).
  
\bibitem{Machacek:1984zw} 
  M.~E.~Machacek and M.~T.~Vaughn,
  ``Two Loop Renormalization Group Equations in a General Quantum Field Theory. 3. Scalar Quartic Couplings,''
  Nucl.\ Phys.\ B {\bf 249}, 70 (1985).

\end{thebibliography}
\end{document}